\documentclass[11pt]{article}
\usepackage[T1]{fontenc}
\usepackage{amsfonts}
\usepackage{amssymb}
\usepackage{amsmath}
\usepackage[dvips]{graphicx,epsfig}
\usepackage{verbatim}
\usepackage{hyperref}
{\catcode`\|=\active 
  \gdef\Braket#1{\left<\mathcode`\|"8000\let|\BraVert {#1}\right>}}
\def\BraVert{\egroup\,\mid@vertical\,\bgroup}
{\catcode`\|=\active
  \gdef\set#1{\mathinner{\lbrace\,{\mathcode`\|"8000\let|\midvert #1}\,\rbrace}}
  \gdef\Set#1{\left\{\:{\mathcode`\|"8000\let|\SetVert #1}\:\right\}}}
\def\midvert{\egroup\mid\bgroup}
\def\SetVert{\egroup\;\mid@vertical\;\bgroup}

\def\slasha#1{\setbox0=\hbox{$#1$}#1\hskip-\wd0\hbox to\wd0{\hss\sl/\/\hss}}
\def\slashb#1{\setbox0=\hbox{$#1$}#1\hskip-\wd0\dimen0=5pt\advance
       \dimen0 by-\ht0\advance\dimen0 by\dp0\lower0.5\dimen0\hbox
         to\wd0{\hss\sl/\/\hss}}

\newcommand{\R}{\mathbb{R}}

\newcommand{\A}{\mathcal{A}}

\newcommand{\Ha}{\mathcal{H}}

\title{\vspace{-2.5cm}\begin{flushright}
 \small SU-4252-869\\
\end{flushright}
\vspace{2.75cm} Twisted Poincaré Invariance, Noncommutative Gauge Theories and UV-IR Mixing}
\author{A. P. Balachandran$^a$\footnote{bal@physics.syr.edu}, A. Pinzul$^b$\footnote{apinzul@fma.if.usp.br}, A. R. Queiroz$^c$\footnote{amilcarq@gmail.com} \\ \\ $^a$ Department of Physics, Syracuse University, \\ Syracuse NY, 13244-1130, USA. \\ $^b$ Insituto de F\'{\i}sica, Universidade de S\~ao Paulo, \\ C.P. 66318, 05315-970, S\~ao Paulo, SP, Brazil. \\ $^c$ Centro Internacional de F\'{\i}sica da Mat\'eria Condensada,\\ Universidade de Bras\'{\i}lia, C.P. 04667, Bras\'{\i}lia, DF, Brazil.}
\begin{document}

\maketitle

%\preprint{SU}
\begin{abstract}
In the absence of gauge fields, quantum field theories on the Groene\-wold-Moyal (GM) plane are invariant under a twisted action of the Poincaré group if they are formulated following \cite{Chaichian:2004za,Aschieri:2005yw,Balachandran:2006ib,Balachandran:2007kv,Balachandran:2007yf,Balachandran:2007vx}. In that formulation, such theories also have no UV-IR mixing \cite{Balachandran:2005pn}. Here we investigate UV-IR mixing in gauge theories with matter following the approach of \cite{Balachandran:2006ib,Balachandran:2007kv}. We prove that there is UV-IR mixing in the one-loop diagram of the $S$-matrix involving a coupling between gauge and matter fields on the GM plane, the gauge field being nonabelian. There is no UV-IR mixing if it is abelian. 
\end{abstract}

\section{Introduction}

A basic result in the study of field theories on the Groenewold-Moyal (GM) plane is the fact that such theories are usually not invariant under the standard action of the Poincaré group, but may be invariant under its twisted action \cite{Chaichian:2004za,Aschieri:2005yw}. This twisted action, which is given by Drinfelds' twisted coproduct, has striking consequences. An adequate understanding of the consequences of such twisted symmetries is yet to be achieved.

There are mainly two distinct approaches to the study of the consequences of twisted Poincaré invariance on the GM plane. One may be represented by works like \cite{Aschieri:2005zs} and the other may be represented by works like \cite{Balachandran:2006ib}. Both approaches start with the algebra of functions $\A_\theta(\R^d)$ on $\R^d$ with the $*$ product defined as
\begin{equation}
 \label{eq:moyal-product}
 f*g(x):=m_\theta(f\otimes g)(x)=m_0(\mathcal{F}_\theta f\otimes g)(x),
\end{equation}
where $m_0(f\otimes g)(x):=f(x)\cdot g(x)$ stands for the usual pointwise multiplication, and
\begin{equation}
  \label{Drinfeld-cocycle}
 \mathcal{F}_\theta=e^{\frac{i}{2}\theta^{\mu \nu}\partial_\mu\otimes\partial_\nu},
\end{equation}
with $\theta^{\mu \nu}=-\theta^{\nu \mu}\equiv \textrm{constant}$, is the Drinfel'd twist element. Both treat matter fields as fields on the GM plane. The difference between the two approaches lies in the treatment of gauge fields. For \cite{Aschieri:2005zs}, gauge fields also are fields on $\A_\theta(\R^d)$, whereas for \cite{Balachandran:2006ib,Balachandran:2007kv}, they are fields based on the commutative algebra $\A_0(\R^d)$.

The present work investigates some properties of the approach due to \cite{Balachandran:2006ib,Balachandran:2007kv} for the construction of field theories on the GM plane.

A key result in the approach of \cite{Balachandran:2007kv} is that it is possible to write noncommutative {\it matter} fields $\phi_\theta\in\A_{\theta}(\R^d)$ in terms of their commutative ($\theta^{\mu\nu}\to 0$) limit $\phi_0\in\A_0(R^d)$ as
\begin{equation}
 \label{eq:dress-transform}
\phi_\theta(x)=\phi_0(x)e^{\frac{1}{2}\overleftarrow{\partial}\wedge P},
\end{equation}
where $\overleftarrow{\partial}\wedge P=\overleftarrow{\partial_\mu}\theta^{\mu \nu} P_\nu$, with $P_\nu$ being the total momentum operator. In contrast, gauge fields are commutative:
\begin{equation}
 A_{\mu,\theta}=A_{\mu,0}.
\end{equation}

Also for \cite{Balachandran:2007kv} the covariant derivative $D_\mu$ acts on noncommutative matter fields $\phi_\theta$ according to
\begin{equation}
 \label{eq:nc-covariant-derivative}
\left(D_\mu\phi\right)_\theta=\left( (D_\mu)_0\phi_0\right)e^{\frac{1}{2}\overleftarrow{\partial}\wedge P},
\end{equation}
where $(D_\mu)_0$ is the standard covariant derivative acting on the commutative field $\phi_0$.

In this work we are going to use the preceding ingredients to analyse one-loop diagrams contributing to quark-gluon vertex.

We prove that there exists a new type of UV-IR mixing in these diagrams. For this proof the existence of the twisted gluon propagator is crucial. Its construction is recalled in section 2. Then in section 3, we examine the one-loop diagram and proceed to single out the integrals in the one-loop diagram that show UV-IR mixing. We close this work with concluding remarks.

Some words of caution before we proceed to the work itself. In the usual construction of field theories on the GM plane, the UV-IR mixing is related to the appearance of nonplanar loop diagrams \cite{Minwalla:1999px}. Such mixing happens also only in the Euclidian formulation of field theories. It does not occur in the Minkowski formulation as clearly shown by Bahns et al. \cite{Bahns:2002vm,Bahns:2004fc}. In the present approach there are no nonplanar loop diagrams and we work with the Minkowski metric. Nevertheless, there is UV-IR mixing according to our criteria: we assume the absence of UV-IR mixing in a loop diagram only if regularization does not mix up UV and IR regions.

An important point to note is that the $S$-matrix for the interaction of nonabelian gauge fields with matter is not Lorentz invariant \cite{Balachandran:2007yf}. This noninvariance is caused by the nonlocality of quantum fields on the GM plane and shows up in scattering amplitudes through their dependence on $\theta^{0i}P_{\textrm{inc.},i}$, $\vec{P}_{\textrm{inc.}}$ being the total incident spatial momentum.

\section{QCD on the GM Plane}

For specificity, we consider QCD. But similar results can be obtained in any nonabelian gauge theory with matter.

Following \cite{Balachandran:2007kv}, the interaction Hamiltonian for QCD on the GM plane can be split into two parts:
\begin{eqnarray}
 \label{eq:int-hamiltonian}
 H_\theta^I&=&\int d^3x ~\Ha^I_\theta, \nonumber\\
 \Ha^I_\theta&=& \Ha^M_\theta+\Ha^G_\theta, 
\end{eqnarray}
where $\Ha_\theta^M$ stands for the matter Hamiltonian density, which contains both quark fields $\psi$ and gluons fields $A_{\mu}$ from quark-gluon interaction, and $\Ha_\theta^G$ stands for the gluon Hamiltonian density, which contains only the gluon fields $A_{\mu}$. We may express these fields for $\theta^{\mu \nu}\neq 0$ in terms of fields for $\theta^{\mu \nu}=0$ as
\begin{eqnarray}
 \label{eq:int-hamiltonian-2}
\Ha^M_\theta&=&\Ha^M_0 e^{\frac{1}{2}\overleftarrow{\partial}\wedge P}, \\
\Ha^G_\theta&=&\Ha^G_0,
\end{eqnarray}
where 
\begin{eqnarray}
\Ha^M_0&=&e\bar{\psi}_0 \gamma^\mu\psi_0 A_{\mu,0}^a\lambda_a, \\
\Ha^G_0&=&g (\partial_\mu A_{\nu,0}^a - \partial_\nu A_{\nu,0}^a)A_{\mu,0}^b A_{\nu,0}^c\lambda_a\lambda_b\lambda_c+O(A^4),
\end{eqnarray}
with $\lambda_a$ being the generators of the Lie algebra of $SU(2)$, are the usual commutative interaction Hamiltonian density terms, with $e$ and $g$ being coupling constants. We will not need $A^4$-order terms, since they are irrelevant for the forthcoming UV-IR mixing analysis.

The $S$-matrix is defined as
\begin{equation}
 \label{eq:S-matrix}
S=T~exp\left( -i\int d^4 x \Ha^I_\theta(x)\right),
\end{equation}
where $\Ha_I(x)=\Ha^M_\theta+\Ha^G_\theta$ is the interaction Hamiltonian density in interaction representation, and $T$ is the time-ordering operator. In second order, the $S$-matrix is written as
\begin{equation}
 \label{eq:S-matrix-2}
S^{(2)}=\frac{(-i)^2}{2!}\int d^4 x d^4 y ~T\left( \Ha^I_\theta(x)\Ha^I_\theta(y)\right),
\end{equation}
where
\begin{equation}
 \label{eq:time-ordered-1}
 T\left( \Ha^I_\theta(x)\Ha^I_\theta(y)\right)=\Theta(x_0-y_0)~\Ha^I_\theta(x)\Ha^I_\theta(y)+\Theta(y_0-x_0)~\Ha^I_\theta(y)\Ha^I_\theta(x),	
\end{equation}
with $\Theta$ being the step function.

It can be shown \cite{Balachandran:2007yf}, that (\ref{eq:S-matrix}) is independent of spatial components of $\theta$, so that we can set them to zero. We do so hereafter.

\section{The Twisted Gluon Propagator}

\begin{figure}[h!]
\begin{center}
\includegraphics[scale=0.5]{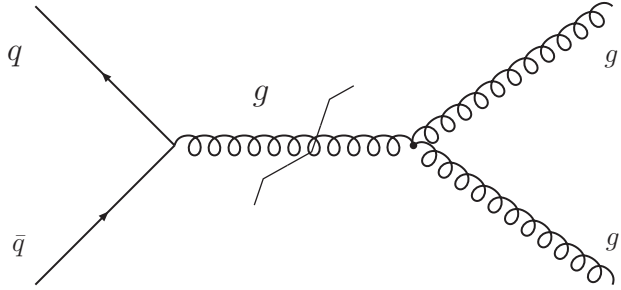}
\caption{The twisted gluon propagator. The twisting of the gluon propagator is symbolized in the figure by a zig-zag line cutting the propagator line. Observe that the external lines are not twisted.}
\label{FIG1}
\end{center}
\end{figure}

The twisted gluon propagator appears as a simplification for the diagram $q\bar{q}\to g g$ and after a change of variables, that shifts the twisting from the $q-\bar{q}-g$ vertex to the $g-g-g$ vertex. For details see \cite{Balachandran:2007yf}. 

The twisted gluon propagator depicted in figure \ref{FIG1} is defined \cite{Balachandran:2007yf} by
\begin{equation}
T\Braket{A^a_{\mu}(x)\left(e^{\frac{1}{2}\overrightarrow{\partial_0}(\theta\cdot P)_0}A^b_{\nu}(y)\right)}=\delta^{ab}\eta_{\mu \nu}D_{F}^\theta(x-y),
\end{equation}
with $a,b$ being gauge group indices, and where in the Lorentz gauge, $D_{F}^\theta$ is just the twisted propagator of a massless scalar field $A$:
\begin{eqnarray}
 \label{eq:gluon-propagator}
&D_F^\theta(x-y)=& \\ &=\nonumber\Theta(x_0-y_0)\Braket{A(x)\left(e^{\frac{1}{2}\overrightarrow{\partial_0}(\theta\cdot P)_0}A(y)\right) }+\Theta(y_0-x_0)\Braket{\left(e^{\frac{1}{2}\overrightarrow{\partial_0}(\theta\cdot P)_0}A(y)\right) A(x)} .
\end{eqnarray}
The first braket in the RHS of (\ref{eq:gluon-propagator}) can be simplified as
\begin{equation}
 \label{eq:braket1}
 \Braket{A(x)\left(e^{\frac{1}{2}\overrightarrow{\partial_0}(\theta\cdot P)_0}A(y)\right)} =\int \frac{d^3 \vec{k}}{(2\pi)^3}\frac{1}{2|\vec{k}|} ~e^{-ik_0(x_0-\frac{1}{2}(\theta\cdot P)_0-y_0)+i\vec{k}\cdot (\vec{x}-\vec{y})}.
\end{equation}
with $(\theta\cdot P)_0=\sum_i \theta^{0i}P_i$. In the same way, the second bracket in the RHS of (\ref{eq:gluon-propagator}) can be simplified as
\begin{equation}
 \label{eq:braket2}
 \Braket{\left(e^{\frac{1}{2}\overrightarrow{\partial_0}(\theta\cdot P)_0}A(y)\right) A(x) }= \int \frac{d^3 \vec{k}}{(2\pi)^3}\frac{1}{2|\vec{k}|} ~e^{-ik_0(y_0+\frac{1}{2}(\theta\cdot P)_0-x_0)+i\vec{k}\cdot (\vec{y}-\vec{x})}.
\end{equation}
Inserting (\ref{eq:braket1}) and (\ref{eq:braket2}) back into (\ref{eq:gluon-propagator}), we obtain
\begin{eqnarray}
 \label{eq:gluon-propagator2}
 D_F^\theta(x,y)=\int \frac{dE}{2\pi i}\frac{d^3\vec{k}}{(2\pi)^3} \frac{1}{2|\vec{k}|}\left(\frac{e^{\frac{i}{2}k_0(\theta\cdot P)_0}}{E-k_0+i\epsilon}-\frac{e^{-\frac{i}{2}k_0(\theta\cdot P)_0}}{E+k_0-i\epsilon} \right)\times \nonumber\\ \times~ e^{-iE(x_0-y_0)+i\vec{k}\cdot(\vec{x}-\vec{y})}.& &
\end{eqnarray}
This expression may be rewritten in a covariant form as
\begin{equation}
 \label{eq:gluon-propagator3}
 D_F^\theta(x,y)=\int \frac{d^4 k}{(2\pi)^4 i}~ \frac{e^{-ik\cdot(x-y)}}{k^2-i\epsilon}~T(k;(\theta\cdot P)_0),% e^{-ik\cdot(x-y)},
\end{equation}
where 
\begin{equation}
 \label{eq:NC-causal-function}
 T(k;(\theta\cdot P)_0)=\cos\left(\frac{|\vec{k}|(\theta\cdot P)_0}{2}\right)+i\frac{k_0}{|\vec{k}|}\sin\left( \frac{|\vec{k}|(\theta\cdot P)_0}{2}\right)
\end{equation}
is the term responsible for the deformation of the gluon propagator due to noncommutativity.

It should be noted that the twisting of the gluon propagator is a reinterpretation of the second order term of the $S$-matrix (\ref{eq:S-matrix-2}). At the beginning, the twisting is in the quark-gluon vertex associated with the term $\mathcal{H}_\theta^M$, which represents the interaction between two quarks and one gluon. In expanding (\ref{eq:S-matrix-2}), and after some algebra we may translate the twisting from the quark-gluon vertex to the gluon propagator line. For detail see \cite{Balachandran:2007yf}.  This procedure simplifies the Feynman rules. In the new Feynman rules, we should distinguish between twisted gluon lines from untwisted ones. Therefore we mark the twisted gluon propagator with a zigzag line cutting the gluon propagator.

\section{One-Loop Diagrams}

\begin{figure}[h!]
\begin{center}
\includegraphics[scale=0.5]{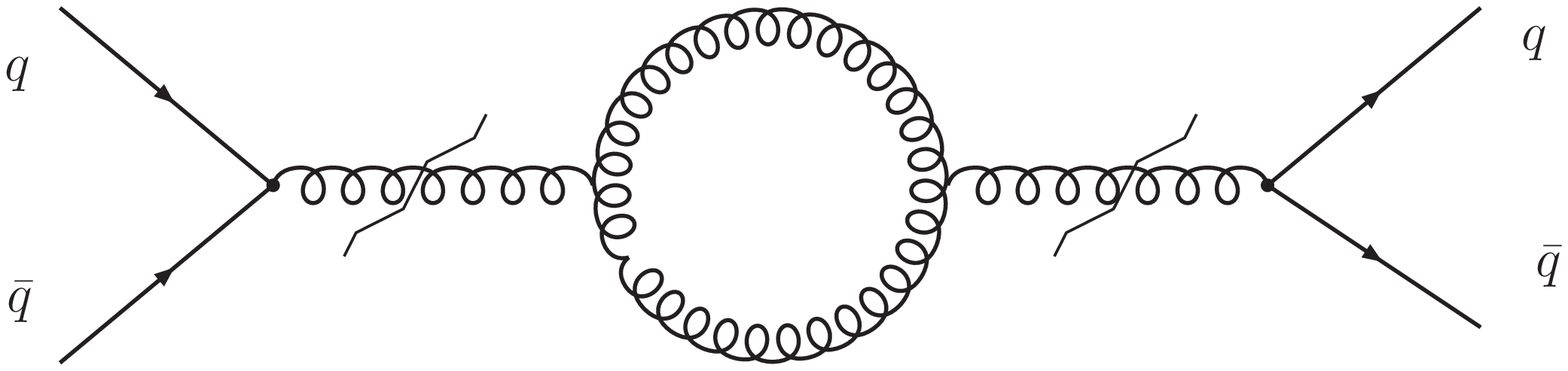}
\caption{A $q\bar{q}\to q\bar{q}$ process containing one loop. The propagators in the gluon loop are not twisted. This kind of diagram does not possess divergence due to twisting.}
\label{FIG2}
\end{center}
\end{figure}

\begin{figure}[h!]
\begin{center}
\includegraphics[scale=0.5]{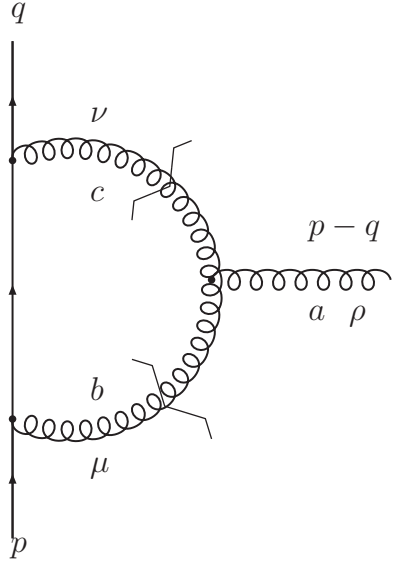}
\caption{A $q\bar{q}\to g$ process containing a loop formed with one quark line and two twisted gluon lines. This kind of propagator leads to UV-IR mixing, which is logarithmic for this diagram.}
\label{FIG3}
\end{center}
\end{figure}

\begin{comment}
\begin{figure}[h!]
\begin{center}
\includegraphics[scale=0.4]{1loop-qqg-bal2.eps}
\caption{A $q\bar{q}\to g$ process containing a loop formed with one quark line and two twisted gluon lines. This kind of propagator leads to UV-IR mixing, which is logarithmic for this diagram.}
\label{FIG3}
\end{center}
\end{figure}
\end{comment}

In this section we analyse one-loop diagrams constructed out of the twisted gluon propagator. There are two possible such one-loop diagrams, with loops containing gluon propagators. The first one is depicted in figure \ref{FIG2} and the second one is depicted in figure \ref{FIG3}.

In the first one-loop diagram (figure \ref{FIG2}), the loop lines are not twisted. Therefore, we do not expect divergence due to twisting in this case. In fact, the absence of such divergences can be seen using momentum conservation. 

In the second one-loop diagram (figure \ref{FIG3}), the loop lines are twisted. As we are going to see below, this twisting of the loop lines leads to an UV-IR mixing logarithmic divergence. 

The loop in figure \ref{FIG3} contains two gluon legs and one quark leg. It may be represented by the following integral:
\begin{equation}
\begin{array}{c}
 \Lambda^{a}_\rho=\textrm{const.}\times g^3~f_{abc}\lambda^{b}\lambda^{c}~\times   \nonumber \\ \vspace{0.3cm}
\times \int\frac{d^4 k}{(2\pi)^4} \frac{\gamma_\mu (\slashb{k}+m) \gamma_\nu [(2p-q-k)_\nu\eta_{\mu \rho}+(2k-p-q)_\rho\eta_{\mu \nu}+(2q-k-p)_\mu\eta_{\nu \rho}]}{k^2+m^2-i\epsilon}\times  \nonumber \\ \vspace{0.3cm}
\times \frac{T(p-k;(\theta \cdot P)_0)T(q-k;(\theta \cdot P)_0)}{[(p-k)^2-i\epsilon][(q-k)^2-i\epsilon]} \nonumber
\end{array}
\end{equation}
We can see that one can split this loop integral in three kinds of integrals with respect to the degree of the integrated variable in the numerator\footnote{We are only interested in the functional form of the integrand with respect to the variable $k$. Thus we do not need to consider the Dirac matrices $\gamma_\mu$, structure constant $f_{abc}$ and so on in the expression of $I(p,q)^n_\theta$. By $k^n$, we mean any monomial of degree $n$ in $k_\mu$.}, i.e., upto constants,
\begin{equation}
I(p,q)_\theta^n=\int \frac{d^4 k}{(2\pi)^4} \frac{k^n}{(k^2+m^2-i\epsilon)}\frac{T(p-k;(\theta \cdot P)_0)T(q-k;(\theta \cdot P)_0)}{[(p-k)^2-i\epsilon][(q-k)^2-i\epsilon]}, \\
\end{equation}
with $n=0,1,2$. Now, dimensional analysis shows that ultraviolet divergent terms come from those with $k^2$ in the numerator, i.e., $I(p,q)_\theta^2$.

In what follows we will be interested only in $I(p,k)_\theta^2$. It can be written as an integral of the form
\begin{eqnarray}
 \label{eq:1loop-1}
I(p,q)^2_\theta&=&\int \frac{d^4 k}{(2\pi)^4} ~\frac{k^2}{(k^2-m^2)}~\frac{T(p-k;(\theta \cdot P)_0)T(q-k;(\theta\cdot P)_0)}{[(p-k)^2-i\epsilon][(q-k)^2-i\epsilon]} = \nonumber \\
  &=& m^2~I(p,q)^0_\theta+\int \frac{d^4 k}{(2\pi)^4}~\frac{T(p-k;(\theta \cdot P)_0)T(q-k;(\theta\cdot P)_0)}{[(p-k)^2-i\epsilon][(q-k)^2-i\epsilon]} \nonumber. 
\end{eqnarray}
The divergent integral is collected in $\mathcal{I}=I(p,q)^2_\theta-m^2~I(p,q)^0_\theta$, i.e.,
\begin{equation}
 \mathcal{I}=\int \frac{d^4 k}{(2\pi)^4}~\frac{T(p-k;(\theta \cdot P)_0)T(q-k;(\theta\cdot P)_0)}{[(p-k)^2-i\epsilon][(q-k)^2-i\epsilon]}. 
\end{equation}
We now as usual rotate $k_0$ integration to go to Euclidian space. Then we make a shift in $k$, the integration variable, to $u=p-k$, so that the integral can be simplified to
\begin{equation}
 \mathcal{I}=\int \frac{d^4 u}{(2\pi)^4}~\frac{T(u;(\theta \cdot P)_0)T(a-u;(\theta\cdot P)_0)}{u^2(a-u)^2},
\end{equation}
where $a=q-p$. This integral can be further simplified using the Feynman parametrization as
\begin{eqnarray}
 \label{eq:1loop-2}
 \mathcal{I}&=&\int \frac{d^4 u}{(2\pi)^4} \int_0^1 dx~\frac{T(u;(\theta \cdot P)_0)T(a-u;(\theta\cdot P)_0)}{\left[ x(a-u)^2+(1-x)u^2 \right]^2} = \nonumber \\
  &=&\int_0^1 dx \int \frac{d^4 u}{(2\pi)^4} \frac{T(u;(\theta \cdot P)_0)T(a-u;(\theta\cdot P)_0)}{\left[ (u-ax)^2+a^2x(1-x) \right]^2}, \nonumber
\end{eqnarray}
where in the last equality we have simplified the denominator and exchanged the order of integration. Now, we can redefine the integration variable, so that
\begin{equation}
 \mathcal{I}=\int_0^1 dx \int \frac{d^4 v}{(2\pi)^4} \frac{T(v+ax;(\theta \cdot P)_0)T(a(1-x)-v;(\theta\cdot P)_0)}{\left[ v^2+c(x) \right]^2},
\end{equation}
where $v=u-ax$ and $c(x)=a^2x(1-x)$. 

Analysing the product $T(v+ax;(\theta \cdot P)_0)T(a(1-x)-v;(\theta\cdot P)_0)$ in the numerator of (\ref{eq:1loop-2}), we note that the term with leading divergence contains a product of cosines, $\cos(\frac{|v+ax|(\theta \cdot P)_0}{2})\cos(\frac{|a(1-x)-v|(\theta \cdot P)_0}{2})$. Furthermore, expanding this cosine product in exponentials we see that a typical divergent term in $\mathcal{I}$ can be written in Euclidian space as
\begin{equation}
\label{eq:1loop-3}
%I_{\textrm{term}}=
\int_0^1 dx \int d\Omega_2  \int dv_0 \int dr \frac{r^2e^{ir(\theta\cdot P)_0+i\frac{a}{2}(2x-1)}}{(r^2+v_0^2+c(x))^2},
\end{equation}
where we are using spherical coordinates for the spatial directions, with $d\Omega_2$ being the solid angle differential. There are three other terms in (\ref{eq:1loop-3}), where $(2x-1)$ is replaced by its negative or $\pm 1$. It is enough for us to consider (\ref{eq:1loop-3}), as the final divergence given in below is seen to be independent of $a$, so that these terms add. This integral may be further rewritten as
\begin{equation}
\label{eq:1loop-4}
% I_{\textrm{term}}=
4\pi \int_0^1 dx \left\{ \frac{d}{dc(x)}~\left[\int dv_0 \int dr  \frac{r^2e^{ir(\theta\cdot P)_0+i\frac{a}{2}(2x-1)}}{r^2+v_0^2+c(x)} \right]\right\}.
\end{equation}
Apart from the derivative with respect to $c(x)$, this integral resembles the integral for UV-IR mixing in \cite{Minwalla:1999px}. We now use the same procedure as in \cite{Minwalla:1999px} to tackle the integral inside the square brackets. So, we write the integrand in (\ref{eq:1loop-4}) using the Schwinger parameters to obtain
\begin{eqnarray}
  \label{eq:1loop-5}
 \int dv_0 \int dr  \frac{r^2e^{ir(\theta\cdot P)_0+i\frac{a}{2}(2x-1)}}{r^2+v_0^2+c(x)}=& & \nonumber\\ =\int_0^\infty d\alpha\int dv_0 \int_0^\infty dr  r^2e^{ir(\theta\cdot P)_0+i\frac{a}{2}(2x-1)-\alpha(r^2+v_0^2+c(x))} . \nonumber
\end{eqnarray}
Performing the integrals in $v_0$ and $r$, and keeping only leading terms in $\theta$ and $a^2$ (or $c(x)$), the expression (\ref{eq:1loop-4}) may be rewritten as
\begin{equation}
\label{eq:1loop-6}
% I_{\textrm{term}}=
\int_0^1 dx \frac{d}{dc(x)} \int_0^\infty \frac{d\alpha}{\alpha^2}e^{-\frac{(\theta\cdot P)_0^2}{4\alpha}-\alpha c(x)}. 
\end{equation}
We may regulate the small $\alpha$ or ultraviolet divergence in the above integral by the insertion of the term $exp(-\frac{1}{\Lambda^2 \alpha})$, so that (\ref{eq:1loop-6}) becomes
\begin{equation}
 \label{eq:1loop-7}
%  I_{\textrm{term}}=
\int_0^1 dx \frac{d}{dc(x)} \int_0^\infty \frac{d\alpha}{\alpha^2}e^{-\frac{1}{\alpha~\Lambda_{\textrm{eff}}^2}-\alpha c(x)}
\end{equation}
where $\frac{1}{\Lambda_{\textrm{eff}}^2}=\frac{(\theta\cdot P)_0^2}{4}+\frac{1}{\Lambda^2}$. Performing the integral in $\alpha$ and then taking the derivative with respect to $c(x)$, we finally find that it behaves like
\begin{equation}
 \label{eq:UV-IR-Log-term}
%  I_{\textrm{term}}\sim
 \ln (\frac{1}{\Lambda_{\textrm{eff}}^2}) = \ln \left(\frac{(\theta\cdot P)_0^2}{4}+\frac{1}{\Lambda^2}\right).
\end{equation}
Observe that when $\Lambda\to\infty$ (UV region), the above term goes like $\ln((\theta\cdot P)_0^2)$. Furthermore if we consider the limit $(\theta\cdot P)_0\to 0$ (IR region) together with the above limit, then this term diverges logarithmically. This is the UV-IR mixing divergence in logarithmic form.

\section{Conclusion}

We have shown in this work that QCD on the GM plane (twisted QCD) presents UV-IR mixing divergence. This corroborates expectations already mentioned in \cite{Balachandran:2006ib,Akofor:2007hk}. In the usual formulation of quantum field theories on the GM plane, the UV-IR mixing is associated with the appearance of nonplanar vertices in the Euclidian formulation. But such UV-IR mixing is absent for Minkowski metrics.

We do not have nonplanar vertices and we work in Minkowski space.
%But such nonplanar vertices are not present in the formulation of quantum field theories due to \cite{Balachandran:2006ib}. 

There are however twisted vertices in QCD processes involving interaction between matter and gauge fields in our approach. In this paper, we have singled out a one-loop diagram containing such a twisted vertex, which shows an UV-IR mixing logarithmic divergence. This is so despite the absence of nonplanar diagrams and use of Minkowski metric.

\section*{Acknowledgment}

ARQ thanks Professor Paulo Teotônio-Sobrinho for the hospitability in the Universidade de S\~ao Paulo, where this work was started. The work of APB is supported in part by DOE under grant number DE-FG02-85ER40231. The work of AP is supported by FAPESP grant number 06/56056-0.

\end{document}